\documentstyle[12pt]{article}\textheight
240mm\textwidth 170mm \hoffset=-1.5cm
\begin{document}
\pagestyle{empty}
\noindent
\hspace*{10.7cm} \vspace{-3mm} TUM-HEP-297/97\\
\hspace*{10.7cm} \vspace{-3mm} KEK Preprint 97-188\\
\hspace*{10.7cm} \vspace{-3mm} KANAZAWA-97-18\\

\hspace*{10.7cm} October 1997

\vspace{0.3cm}

\begin{center}
{\Large\bf  
  Sum rules in the superpartner spectrum\\
of the minimal supersymmetric standard model}
\end{center} 

\vspace{1cm}

\begin{center}
{\sc Yoshiharu Kawamura}$\ ^{\dag,*}$\\ 
{\em  Physik Department,  \vspace{-2mm}
Technische Universit\" at M\" unchen \\
 D-85747  \vspace{-2mm} Garching, Germany}\\
{\sc Tatsuo Kobayashi}\\
{\em  Institute of Particle and \vspace{-2mm}
 Nuclear Studies\\
High Energy Accelerator Research  Organization,
 Tanashi,  \vspace{-2mm} Tokyo 188, Japan}\\ 
{\sc Jisuke Kubo}$\ ^{**}$\\
{\em Physics Department, \vspace{-2mm} Faculty of Science, \\
Kanazawa \vspace{-2mm} University,
Kanazawa, 920-11 Japan } 
\end{center}

\vspace{1cm}
\begin{center}
{\sc\large Abstract}
\end{center}
Assuming that the string inspired, universal sum rules for soft
supersymmetry-breaking  terms, which have been recently found both
in a wide class of four-dimensional superstrings and in
supersymmertic gauge-Yukawa unified gauge models, 
are satisfied above and at the grand unification scale,
 we investigate their
low energy consequences and derive
 sum rules in the superpartner spectrum
of the minimal supersymmetric standard model.
\noindent

\vspace*{3cm}
\footnoterule
\vspace*{2mm}
\noindent
$^{\dag}$ Humboldt Fellow. \vspace{-3mm}\\
$^{*}$
On leave from: Department of Physics, Shinshu University,
Matsumoto,  \vspace{-3mm} 390 Japan. \\
$^{**}$Partially supported  by the Grants-in-Aid
for \vspace{-3mm} Scientific Research  from the Ministry of
Education, Science 
and Culture \vspace{-3mm}  (No. 40211213).\\

\newpage
\pagestyle{plain}
One of the most important issues in realistic 
supersymmetric theories
is to understand how  supersymmetry is broken
and then to relate it with low energy physics.
Though the recent exiting theoretical developments in
supersymmetric gauge theories as well as 
superstrings \cite{lerche1},
this problem has not  been solved 
in a satisfactory fashion yet.
It is however widely accepted that the supersymmetry breaking,
whatever its origin is,
appears as  soft supersymmetry-breaking (SSB)
terms in low energy effective theories,
because the softness is a desirable 
property for not spoiling the supersymmetric solution
of the naturalness problem of
 the standard model \cite{nilles1}.

One might hope that the SSB terms have a
minimal structure as it is
suggested by $N=1$ minimal supergravity \cite{nilles1}, on one hand.
It may be worthwhile, on the other hand, to find out
relations among the SSB terms
that have least dependence of the mechanism
of supersymmetry breaking and are satisfied  in a wide class
of models. Phenomenological investigations and
consequences based on
these relations certainly would have a more general validity than
those based on the assumption of the so-called universal SSB terms.
At first sight one might think that once we deviate 
from the universality
of the SSB terms, we would fall into the
 chaos of varieties  \cite{susy}.
Recent investigations on the SSB terms in $4D$ 
superstrings \cite{IL}-\cite{BIM2}, however, 
have shown that 
it is possible to do  systematic investigations
of the SSB terms, and it has turned out to be
also possible to
parametrize the SSB terms in 
a simple way \cite{BIM}--\cite{BIM2} 
so that one can easily find relations among them
that are independent of the detailed nature of supersymmetry
breaking. It has been then shown  \cite{kawamura1} that
there exist symmetries in the effective
$N=1$ supergravity, which, along with
a simple assumption on Yukawa couplings, lead to  these relations.
These symmetries happen to coincide with 
the S- and T- dualities, which are usually present in 
$4D$ superstrings \cite{foot3}.
Moreover it has turned out \cite{kawamura1} that  these relations are
renormalization group (RG) invariant at the lowest nontrivial order
in perturbation theory  in all 
gauge-Yukawa unified (GYU) models \cite{kmz1}, so that
once they are satisfied at the string scale, 
they are satisfied  at the
grand unification  scale, too. 
We call these relations sum rules for the SSB terms,
which may be summarized as \cite{BIM2,kawamura1,kobayashi1}
\begin{eqnarray}
h^{ijk} &=& -M\,Y^{ijk},
\label{sum1}\\
M^2 &=& m^{2}_{i}+m^{2}_{j}+m^{2}_{k}.
\label{sum2}
\end{eqnarray}
Here $M$ and  $m^{2}_{i}$
 stand for the unified gaugino mass and
the soft scalar mass squared of the chiral superfield $\Phi_i$,
respectively. $Y^{ijk}$ is the dimensionless Yukawa coupling
for the $\Phi_i \,\Phi_j \,\Phi_k$ term in the superpotential, 
while the $h^{ijk}$ is
the dimensional coupling
for the trilinear term of the corresponding scalar components.
Higher order corrections to the above sum rules 
are model-dependent in general.
We, furthermore, would like to recall that the assumption
on the gauge-Yukawa unification in supersymmetric grand
unified theories (GUTs), especially in the
third generation sector, leads to a successful prediction of the
top quark mass \cite{kmoz2}.

In this letter we are motivated by the desire to find out
low energy consequences  of the sum rules (\ref{sum1}) and
(\ref{sum2}) which are assumed to be satisfied 
 for the third generation sector of  $SU(5)$ type GUTs at
the GUT scale $M_{\rm GUT}$.
We will assume that between $M_{\rm GUT} (\sim 10^{16}
~\mbox{GeV})$ 
and the supersymmetry braking scale  
$M_{S} ( < 1~\mbox{TeV})$ the minimal supersymmetric
standard model (MSSM)  describes  particle physics,
and we will derive sum rules in the superpartner
spectrum (Eq. (\ref{sum})) from the string inspired,
universal  relations
(\ref{sum1}) and (\ref{sum2}).
These sum rules are independent on the details of
the SSB parameters as long as the 
 sum rules (\ref{sum1}) and
(\ref{sum2}) are satisfied at  $M_{\rm GUT}$.
Needless to say that these sum rules could
be tested by future experiments, e.g., at LHC. 

Before we present the details of our investigations,
we would like to briefly outline the basic nature
that leads to the sum rules (\ref{sum1}) and (\ref{sum2}),
both in $4D$-superstring-inspired supergravity models
and GYU models.

To analyze how the sum rules
(\ref{sum1}) and (\ref{sum2}) within the framework
of  effective $N=1$ supergravity can be realized,
one considers   a 
non-canonical K\"ahler potential of the general form
\begin{eqnarray}
K &=& \tilde{K}(\Phi_a, \Phi^{*a}) + \sum_i K_i^i(\Phi_a, \Phi^{*a})
|\Phi^i|^2,
\end{eqnarray}
where $\Phi_a$'s and $\Phi_i$'s are 
chiral  superfields in the  hidden 
and visible sectors, 
respectively. 
The basic   assumptions are:
(1) Supersymmetry is broken by the $F$-term condensations 
($\langle F_a \rangle \neq 0$) of the   hidden sector fields
 $\Phi_a$.
(2) The gaugino mass $M$ stems from
the gauge kinetic
function $f$ which depends  only on 
the hidden sector fields, i.e. $f=f(\Phi_a)$.
(3) We consider only
those Yukawa couplings that have no field dependence.
(4) The vacuum energy $V_0$ vanishes.
 For the sum rules (\ref{sum1}) and (\ref{sum2}) to be satisfied under
 these assumptions,  a certain relation
among the K\" ahler potential $\tilde{K}$  in the hidden sector, 
the gauge kinetic function $f$ and the K\"ahler metric
has to exist, i.e.,
\begin{eqnarray}
K_{(T)}(\Phi_a,\Phi^{*a}) &\equiv&
\ln (K_i^i K_j^j K_k^k)
=\tilde{K} + \ln \mbox{Re} f 
+\mbox{const.}
\end{eqnarray}
for  all  $\{ i,j,k \}$ appearing in
 the sum rules (\ref{sum1}) and (\ref{sum2}), 
implying that
 the theory has two types of symmetries:
The first one corresponds to  
the K\"ahler transformation together with  the chiral rotation of
the matter multiplets, 
\begin{eqnarray}
\Phi_i & \to &  e^{{\cal M}_i} \Phi_i ~,~
 \Phi^{*i} \to e^{\overline{\cal M}_i} \Phi^{*i}~,~
K_i^i \to  K_i^i  e^{-({\cal M}_i+\overline{\cal M}_i)}
\\
K_{(T)} &\to & K_{(T)} -{\cal M}-\overline{\cal M}~,~
f(\Phi_a) \to  f(\Phi_a)~,~
W \to e^{{\cal M}}\, W,
\end{eqnarray}
where ${\cal M}_i$ is a function of $\Phi_a$ and has to satisfy
the constraint
${\cal M}_i+
{\cal M}_j+{\cal M}_k={\cal M}$ for all possible set of $\{i,j,k \}$
appearing in the sum rules (\ref{sum1}) and (\ref{sum2}).
The second one is the invariance of
the K\"ahler metric $K_{(S)\,b}^{a}$ under 
the $SL(2,R)$ transformation of
the gauge kinetic function
$f(\Phi_a)$, where $
K_{(S)}=-\ln (f(\Phi_a)+\overline{f}(\Phi^{*a}))$.
For 4D string models,
these symmetries
appear as  the target-space duality 
invariance
and $S$-duality \cite{foot3}, respectively.
In fact, Brignole {\em et. al.}
 \cite{BIM2} have already found these sum rules  in
their explicit computations
 in various orbifold models.
In case that gauge symmetries break, we generally have $D$-term contributions 
to the soft scalar masses.
Such $D$-term contributions, however, do not appear in the sum rules,
because each $D$-term contribution is proportional to the charge of 
the matter field $\Phi_i$ \cite{Dterm}.

The basic assumption in GYU models is that 
 the Yukawa couplings $Y^{ijk}$  are 
expressed in terms  of the gauge coupling $g$:
\begin{eqnarray} 
Y^{ijk} &=& \rho^{ijk}g+\dots,
\label{Yg}
\end{eqnarray}
where $\rho^{ijk}$ are constant independent of
$g$ and   $\dots$ stands for higher order terms.
Eq. (\ref{Yg}) is the power series solution to the reduction 
equation \cite{zim1} 
$\beta_{Y}^{ijk} = \beta_{g}\,d Y^{ijk}/d g$,
where $\beta_{Y}^{ijk}$ and $\beta_{g}$ stand for
the $\beta$ functions of $Y^{ijk}$ and $g$, 
respectively.
The next assumption is that 
the coefficients $\rho^{ijk}$ 
satisfy the diagonality relation
$\rho_{ipq}\rho^{jpq} \propto  \delta_{i}^{j}$.
This  implies that the one-loop 
anomalous dimensions $
\gamma_{i}^{(1)\,j}/16 \pi^2$ for $\Phi_i$'s
become diagonal if the reduction solution (\ref{Yg})
 is inserted, i.e., $
\gamma_{i}^{(1)\,j} =
\gamma_{i}\,\delta^{j}_{i}\,g^2$,
where $ \gamma_{i}$ are constant independent of
$g$.
It can be then shown that 
the sum rule (\ref{sum1}) as well as the relation
$(m^2)^{j}_{i} = m^{2}_{i}\,\delta^{j}_{i}=
\kappa_{i} M^2 \delta^{j}_{i}$ with $\kappa_i=
\gamma_i/(T(R)-3 C(G))$
are RG invariant in 
one-loop order,
where $T(R)$ and $C(G)$ are   
the Dynkin index of the  matter representation $R$
and the quadratic Casimir of the adjoint representation 
of $G$, respectively.
 The sum rule (\ref{sum2}) then  follows
from the consequence of the reduction of $Y^{ijk}$, i.e.,
$\gamma_{i}+\gamma_{j}+\gamma_{k} = 
T(R)-3C(G)$
for $\{ i,j,k \}$ appearing in the sum rule.

To come to our main result
of this letter,
let us first describe the parameter space.
Since we assume an $SU(5)$ type GYU in the third generation,
 Eq. (\ref{Yg}) takes the form
\begin{eqnarray}
g_t &=& \rho_t\, g~,~g_b =g_{\tau}=
 \rho_b\, g,
\end{eqnarray}
at $M_{\rm GUT}$, where $g_i~(i=t,b,\tau)$ are
the Yukawa couplings for the top, bottom quarks and the tau, and
we ignore the Cabibbo-Kabayashi-Maskawa mixing of the quarks.
For a given model, the $\rho$'s are fixed, but here we consider
them as free parameters.
It is more convenient to go from the
parameter space $(\rho_t~,~\rho_b)$ to 
another  one $(k_t \equiv \rho_t^2~,~\tan\beta)$, because we use 
 the (physical) top quark mass $M_t$ as input,
i.e., $M_t=(175.6\pm 5.5)$ GeV.
Therefore, the unification condition of the gauge couplings of the MSSM,
along with 
$\alpha^{-1}_{\rm EM}(M_Z)=127.9+
(8/9\pi)\ln (M_t /M_Z)$ and also the tau mass
$M_{\tau}=1.777$ GeV as low energy input parameters, 
fixes the allowed region in the $k_t - \tan\beta$ space,
which is shown in Fig.~1,
where we have used $M_S=300$ GeV.
In the following analyses when varying $\tan\beta$, 
we use $k_t$ for $M_t=175$GeV, while
ignoring the bottom quark mass.

\begin{center}
\setlength{\unitlength}{0.240900pt}
\ifx\plotpoint\undefined\newsavebox{\plotpoint}\fi
\sbox{\plotpoint}{\rule[-0.200pt]{0.400pt}{0.400pt}}%
\begin{picture}(1500,900)(0,0)
\font\gnuplot=cmr10 at 10pt
\gnuplot
\sbox{\plotpoint}{\rule[-0.200pt]{0.400pt}{0.400pt}}%
\put(220.0,113.0){\rule[-0.200pt]{292.934pt}{0.400pt}}
\put(220.0,113.0){\rule[-0.200pt]{0.400pt}{184.048pt}}
\put(220.0,113.0){\rule[-0.200pt]{4.818pt}{0.400pt}}
\put(198,113){\makebox(0,0)[r]{0}}
\put(1416.0,113.0){\rule[-0.200pt]{4.818pt}{0.400pt}}
\put(220.0,189.0){\rule[-0.200pt]{4.818pt}{0.400pt}}
\put(198,189){\makebox(0,0)[r]{0.2}}
\put(1416.0,189.0){\rule[-0.200pt]{4.818pt}{0.400pt}}
\put(220.0,266.0){\rule[-0.200pt]{4.818pt}{0.400pt}}
\put(198,266){\makebox(0,0)[r]{0.4}}
\put(1416.0,266.0){\rule[-0.200pt]{4.818pt}{0.400pt}}
\put(220.0,342.0){\rule[-0.200pt]{4.818pt}{0.400pt}}
\put(198,342){\makebox(0,0)[r]{0.6}}
\put(1416.0,342.0){\rule[-0.200pt]{4.818pt}{0.400pt}}
\put(220.0,419.0){\rule[-0.200pt]{4.818pt}{0.400pt}}
\put(198,419){\makebox(0,0)[r]{0.8}}
\put(1416.0,419.0){\rule[-0.200pt]{4.818pt}{0.400pt}}
\put(220.0,495.0){\rule[-0.200pt]{4.818pt}{0.400pt}}
\put(198,495){\makebox(0,0)[r]{1}}
\put(1416.0,495.0){\rule[-0.200pt]{4.818pt}{0.400pt}}
\put(220.0,571.0){\rule[-0.200pt]{4.818pt}{0.400pt}}
\put(198,571){\makebox(0,0)[r]{1.2}}
\put(1416.0,571.0){\rule[-0.200pt]{4.818pt}{0.400pt}}
\put(220.0,648.0){\rule[-0.200pt]{4.818pt}{0.400pt}}
\put(198,648){\makebox(0,0)[r]{1.4}}
\put(1416.0,648.0){\rule[-0.200pt]{4.818pt}{0.400pt}}
\put(220.0,724.0){\rule[-0.200pt]{4.818pt}{0.400pt}}
\put(198,724){\makebox(0,0)[r]{1.6}}
\put(1416.0,724.0){\rule[-0.200pt]{4.818pt}{0.400pt}}
\put(220.0,801.0){\rule[-0.200pt]{4.818pt}{0.400pt}}
\put(198,801){\makebox(0,0)[r]{1.8}}
\put(1416.0,801.0){\rule[-0.200pt]{4.818pt}{0.400pt}}
\put(220.0,877.0){\rule[-0.200pt]{4.818pt}{0.400pt}}
\put(198,877){\makebox(0,0)[r]{2}}
\put(1416.0,877.0){\rule[-0.200pt]{4.818pt}{0.400pt}}
\put(220.0,113.0){\rule[-0.200pt]{0.400pt}{4.818pt}}
\put(220,68){\makebox(0,0){0}}
\put(220.0,857.0){\rule[-0.200pt]{0.400pt}{4.818pt}}
\put(423.0,113.0){\rule[-0.200pt]{0.400pt}{4.818pt}}
\put(423,68){\makebox(0,0){10}}
\put(423.0,857.0){\rule[-0.200pt]{0.400pt}{4.818pt}}
\put(625.0,113.0){\rule[-0.200pt]{0.400pt}{4.818pt}}
\put(625,68){\makebox(0,0){20}}
\put(625.0,857.0){\rule[-0.200pt]{0.400pt}{4.818pt}}
\put(828.0,113.0){\rule[-0.200pt]{0.400pt}{4.818pt}}
\put(828,68){\makebox(0,0){30}}
\put(828.0,857.0){\rule[-0.200pt]{0.400pt}{4.818pt}}
\put(1031.0,113.0){\rule[-0.200pt]{0.400pt}{4.818pt}}
\put(1031,68){\makebox(0,0){40}}
\put(1031.0,857.0){\rule[-0.200pt]{0.400pt}{4.818pt}}
\put(1233.0,113.0){\rule[-0.200pt]{0.400pt}{4.818pt}}
\put(1233,68){\makebox(0,0){50}}
\put(1233.0,857.0){\rule[-0.200pt]{0.400pt}{4.818pt}}
\put(1436.0,113.0){\rule[-0.200pt]{0.400pt}{4.818pt}}
\put(1436,68){\makebox(0,0){60}}
\put(1436.0,857.0){\rule[-0.200pt]{0.400pt}{4.818pt}}
\put(220.0,113.0){\rule[-0.200pt]{292.934pt}{0.400pt}}
\put(1436.0,113.0){\rule[-0.200pt]{0.400pt}{184.048pt}}
\put(220.0,877.0){\rule[-0.200pt]{292.934pt}{0.400pt}}
\put(45,495){\makebox(0,0){$k_t$}}
\put(828,23){\makebox(0,0){$\tan \beta$}}
\put(828,495){\makebox(0,0)[r]{$M_t=180$ [GeV]}}
\put(828,266){\makebox(0,0)[r]{$M_t=170$ [GeV]}}
\put(220.0,113.0){\rule[-0.200pt]{0.400pt}{184.048pt}}
\put(261,773){\usebox{\plotpoint}}
\multiput(261.58,743.69)(0.496,-8.853){37}{\rule{0.119pt}{7.060pt}}
\multiput(260.17,758.35)(20.000,-333.347){2}{\rule{0.400pt}{3.530pt}}
\multiput(281.58,419.94)(0.496,-1.415){37}{\rule{0.119pt}{1.220pt}}
\multiput(280.17,422.47)(20.000,-53.468){2}{\rule{0.400pt}{0.610pt}}
\multiput(301.58,366.84)(0.496,-0.523){37}{\rule{0.119pt}{0.520pt}}
\multiput(300.17,367.92)(20.000,-19.921){2}{\rule{0.400pt}{0.260pt}}
\multiput(321.00,346.92)(1.069,-0.491){17}{\rule{0.940pt}{0.118pt}}
\multiput(321.00,347.17)(19.049,-10.000){2}{\rule{0.470pt}{0.400pt}}
\multiput(342.00,336.93)(2.157,-0.477){7}{\rule{1.700pt}{0.115pt}}
\multiput(342.00,337.17)(16.472,-5.000){2}{\rule{0.850pt}{0.400pt}}
\multiput(362.00,331.95)(4.258,-0.447){3}{\rule{2.767pt}{0.108pt}}
\multiput(362.00,332.17)(14.258,-3.000){2}{\rule{1.383pt}{0.400pt}}
\put(382,328.17){\rule{4.100pt}{0.400pt}}
\multiput(382.00,329.17)(11.490,-2.000){2}{\rule{2.050pt}{0.400pt}}
\put(402,326.17){\rule{4.300pt}{0.400pt}}
\multiput(402.00,327.17)(12.075,-2.000){2}{\rule{2.150pt}{0.400pt}}
\multiput(524.00,326.60)(14.665,0.468){5}{\rule{10.200pt}{0.113pt}}
\multiput(524.00,325.17)(79.829,4.000){2}{\rule{5.100pt}{0.400pt}}
\multiput(625.00,330.59)(7.739,0.485){11}{\rule{5.929pt}{0.117pt}}
\multiput(625.00,329.17)(89.695,7.000){2}{\rule{2.964pt}{0.400pt}}
\multiput(727.00,337.58)(5.238,0.491){17}{\rule{4.140pt}{0.118pt}}
\multiput(727.00,336.17)(92.407,10.000){2}{\rule{2.070pt}{0.400pt}}
\multiput(828.00,347.58)(3.692,0.494){25}{\rule{2.986pt}{0.119pt}}
\multiput(828.00,346.17)(94.803,14.000){2}{\rule{1.493pt}{0.400pt}}
\multiput(929.00,361.58)(3.054,0.495){31}{\rule{2.500pt}{0.119pt}}
\multiput(929.00,360.17)(96.811,17.000){2}{\rule{1.250pt}{0.400pt}}
\multiput(1031.00,378.58)(2.041,0.497){47}{\rule{1.716pt}{0.120pt}}
\multiput(1031.00,377.17)(97.438,25.000){2}{\rule{0.858pt}{0.400pt}}
\multiput(1132.00,403.58)(1.451,0.498){67}{\rule{1.254pt}{0.120pt}}
\multiput(1132.00,402.17)(98.397,35.000){2}{\rule{0.627pt}{0.400pt}}
\multiput(1233.00,438.58)(1.088,0.495){35}{\rule{0.963pt}{0.119pt}}
\multiput(1233.00,437.17)(39.001,19.000){2}{\rule{0.482pt}{0.400pt}}
\multiput(1274.00,457.58)(0.837,0.496){45}{\rule{0.767pt}{0.120pt}}
\multiput(1274.00,456.17)(38.409,24.000){2}{\rule{0.383pt}{0.400pt}}
\multiput(1314.00,481.58)(0.754,0.494){25}{\rule{0.700pt}{0.119pt}}
\multiput(1314.00,480.17)(19.547,14.000){2}{\rule{0.350pt}{0.400pt}}
\multiput(1335.00,495.58)(0.625,0.494){29}{\rule{0.600pt}{0.119pt}}
\multiput(1335.00,494.17)(18.755,16.000){2}{\rule{0.300pt}{0.400pt}}
\multiput(1355.58,511.00)(0.498,0.512){77}{\rule{0.120pt}{0.510pt}}
\multiput(1354.17,511.00)(40.000,39.941){2}{\rule{0.400pt}{0.255pt}}
\multiput(1395.58,552.00)(0.498,0.757){79}{\rule{0.120pt}{0.705pt}}
\multiput(1394.17,552.00)(41.000,60.537){2}{\rule{0.400pt}{0.352pt}}
\put(423.0,326.0){\rule[-0.200pt]{24.331pt}{0.400pt}}
\sbox{\plotpoint}{\rule[-0.500pt]{1.000pt}{1.000pt}}%
\put(261,605){\usebox{\plotpoint}}
\multiput(261,605)(1.636,-20.691){13}{\usebox{\plotpoint}}
\multiput(281,352)(10.080,-18.144){2}{\usebox{\plotpoint}}
\put(314.40,306.62){\usebox{\plotpoint}}
\put(333.05,297.98){\usebox{\plotpoint}}
\put(353.10,292.78){\usebox{\plotpoint}}
\put(373.62,289.84){\usebox{\plotpoint}}
\put(394.32,288.38){\usebox{\plotpoint}}
\put(415.05,287.38){\usebox{\plotpoint}}
\multiput(423,287)(20.756,0.000){5}{\usebox{\plotpoint}}
\multiput(524,287)(20.746,0.616){5}{\usebox{\plotpoint}}
\multiput(625,290)(20.731,1.016){5}{\usebox{\plotpoint}}
\multiput(727,295)(20.706,1.435){4}{\usebox{\plotpoint}}
\multiput(828,302)(20.655,2.045){5}{\usebox{\plotpoint}}
\multiput(929,312)(20.589,2.624){5}{\usebox{\plotpoint}}
\multiput(1031,325)(20.434,3.642){5}{\usebox{\plotpoint}}
\multiput(1132,343)(20.100,5.174){5}{\usebox{\plotpoint}}
\multiput(1233,369)(19.785,6.273){2}{\usebox{\plotpoint}}
\multiput(1274,382)(19.102,8.118){2}{\usebox{\plotpoint}}
\multiput(1314,399)(18.739,8.923){2}{\usebox{\plotpoint}}
\put(1352.54,419.52){\usebox{\plotpoint}}
\multiput(1355,421)(16.804,12.183){2}{\usebox{\plotpoint}}
\multiput(1395,450)(13.979,15.342){3}{\usebox{\plotpoint}}
\put(1436,495){\usebox{\plotpoint}}
\multiput(274,877)(0.442,-20.751){16}{\usebox{\plotpoint}}
\multiput(281,548)(4.191,-20.328){5}{\usebox{\plotpoint}}
\multiput(301,451)(10.523,-17.890){2}{\usebox{\plotpoint}}
\put(328.80,411.06){\usebox{\plotpoint}}
\put(345.87,399.45){\usebox{\plotpoint}}
\put(365.28,392.18){\usebox{\plotpoint}}
\put(385.48,387.48){\usebox{\plotpoint}}
\put(406.03,384.62){\usebox{\plotpoint}}
\multiput(423,383)(20.754,-0.205){5}{\usebox{\plotpoint}}
\multiput(524,382)(20.719,1.231){5}{\usebox{\plotpoint}}
\multiput(625,388)(20.656,2.025){5}{\usebox{\plotpoint}}
\multiput(727,398)(20.559,2.850){5}{\usebox{\plotpoint}}
\multiput(828,412)(20.398,3.837){5}{\usebox{\plotpoint}}
\multiput(929,431)(20.112,5.127){5}{\usebox{\plotpoint}}
\multiput(1031,457)(19.551,6.969){5}{\usebox{\plotpoint}}
\multiput(1132,493)(18.379,9.644){6}{\usebox{\plotpoint}}
\multiput(1233,546)(16.945,11.986){2}{\usebox{\plotpoint}}
\multiput(1274,575)(15.427,13.885){3}{\usebox{\plotpoint}}
\put(1327.06,624.06){\usebox{\plotpoint}}
\multiput(1335,632)(13.287,15.945){2}{\usebox{\plotpoint}}
\multiput(1355,656)(11.125,17.522){3}{\usebox{\plotpoint}}
\multiput(1395,719)(8.373,18.992){5}{\usebox{\plotpoint}}
\put(1436,812){\usebox{\plotpoint}}
\end{picture}

FIG. 1: $M_t$ in the $k_t - \tan \beta$ space
\end{center}

\noindent
The parameter space in the SSB sector 
is  constrained at  $M_{\rm GUT}$ due
to unification:
\begin{eqnarray}
h_t &=&-M \,g_t=-M \rho_t \,g~,
~h_b =h_{\tau}=-M\, g_b=-M \rho_b\, g,\\
m_{\tilde{t}_R}^{2} &=&m_{\tilde{t}_L}^{2}=
m_{\tilde{b}_L}^{2}= m_{\tilde{\tau}_R}^{2}~,~
m_{\tilde{b}_R}^{2}= m_{\tilde{\tau}_L}^{2},
\end{eqnarray}
and we define
\begin{eqnarray}
m_{\Sigma (t)}^{2}&\equiv& m_{\tilde{t}_R}^{2}+
m_{\tilde{t}_L}^{2}+m_{H_2}^{2}~,~
m_{\Sigma (b,\tau)}^{2}\equiv 
m_{\tilde{b}_R,\tilde{\tau}_R}^{2}+
m_{\tilde{b}_L,\tilde{\tau}_L}^{2}+m_{H_1}^{2}.
\end{eqnarray}
The unification condition for the gaugino mass is:
$M_1=M_2=M_3=M$ at $M_{\rm GUT}$. 
We note that in the one-loop RG evolution of
$m_{\Sigma}^{2}$'s 
in the MSSM only  the same combinations 
of the sum of $m_{i}^{2}$'s
enter. Therefore, as far as we are concerned with
the evolution of $m_{\Sigma}^{2}$'s,
we have only one additional parameter
$M$, which we further identify with $M_{S}$.
In Fig.~2, we show the evolution of 
$m_{\Sigma(t,b,\tau)}^{2}/M_{3}^2$ 
 as function of $Q/M_S$
for $\tan\beta =50$.

\begin{center}
\input{m2.tex}

FIG. 2: The evolution of
$m^2_\Sigma/M_3^2$ for $\tan \beta =50$.
\end{center}

\noindent
The figure above shows that $m^2_{\Sigma(t,b)}/M_3^2$ is
 stable
near $\ln Q/M_S \simeq 0$.
As for $m^2_{\Sigma(\tau)}/M_3^2$, we find
$\Delta(m^2_{\Sigma(\tau)}/M_3^2) \simeq  \pm 0.1$
for $\Delta(\ln Q/M_S) \simeq \pm 5$ at $\tan\beta =50$.
These features in the stability of the evolution of 
$m^2_{\Sigma(t,b,\tau)}/M_3^2$ do not change very much
for the entire range of $\tan\beta$ between $2$ and $50$.

To derive the announced sum rules for the superpartner spectrum,
we further define
\begin{eqnarray}
s_i &\equiv& m^2_{\Sigma(i)}/M_3^2~~(i=t,b,\tau)~~\mbox{at}~~
Q=M_S.
\label{sbt}
\end{eqnarray}
These parameters do not depend on the value of $M$
(which is defined at $M_{\rm GUT}$) in one-loop order,
but they do on $\tan \beta$.
This dependence is shown in Fig.~3.

\begin{center}
\setlength{\unitlength}{0.240900pt}
\ifx\plotpoint\undefined\newsavebox{\plotpoint}\fi
\sbox{\plotpoint}{\rule[-0.200pt]{0.400pt}{0.400pt}}%
\begin{picture}(1500,900)(0,0)
\font\gnuplot=cmr10 at 10pt
\gnuplot
\sbox{\plotpoint}{\rule[-0.200pt]{0.400pt}{0.400pt}}%
\put(220.0,189.0){\rule[-0.200pt]{292.934pt}{0.400pt}}
\put(220.0,113.0){\rule[-0.200pt]{0.400pt}{184.048pt}}
\put(220.0,113.0){\rule[-0.200pt]{4.818pt}{0.400pt}}
\put(198,113){\makebox(0,0)[r]{-0.2}}
\put(1416.0,113.0){\rule[-0.200pt]{4.818pt}{0.400pt}}
\put(220.0,189.0){\rule[-0.200pt]{4.818pt}{0.400pt}}
\put(198,189){\makebox(0,0)[r]{0}}
\put(1416.0,189.0){\rule[-0.200pt]{4.818pt}{0.400pt}}
\put(220.0,266.0){\rule[-0.200pt]{4.818pt}{0.400pt}}
\put(198,266){\makebox(0,0)[r]{0.2}}
\put(1416.0,266.0){\rule[-0.200pt]{4.818pt}{0.400pt}}
\put(220.0,342.0){\rule[-0.200pt]{4.818pt}{0.400pt}}
\put(198,342){\makebox(0,0)[r]{0.4}}
\put(1416.0,342.0){\rule[-0.200pt]{4.818pt}{0.400pt}}
\put(220.0,419.0){\rule[-0.200pt]{4.818pt}{0.400pt}}
\put(198,419){\makebox(0,0)[r]{0.6}}
\put(1416.0,419.0){\rule[-0.200pt]{4.818pt}{0.400pt}}
\put(220.0,495.0){\rule[-0.200pt]{4.818pt}{0.400pt}}
\put(198,495){\makebox(0,0)[r]{0.8}}
\put(1416.0,495.0){\rule[-0.200pt]{4.818pt}{0.400pt}}
\put(220.0,571.0){\rule[-0.200pt]{4.818pt}{0.400pt}}
\put(198,571){\makebox(0,0)[r]{1}}
\put(1416.0,571.0){\rule[-0.200pt]{4.818pt}{0.400pt}}
\put(220.0,648.0){\rule[-0.200pt]{4.818pt}{0.400pt}}
\put(198,648){\makebox(0,0)[r]{1.2}}
\put(1416.0,648.0){\rule[-0.200pt]{4.818pt}{0.400pt}}
\put(220.0,724.0){\rule[-0.200pt]{4.818pt}{0.400pt}}
\put(198,724){\makebox(0,0)[r]{1.4}}
\put(1416.0,724.0){\rule[-0.200pt]{4.818pt}{0.400pt}}
\put(220.0,801.0){\rule[-0.200pt]{4.818pt}{0.400pt}}
\put(198,801){\makebox(0,0)[r]{1.6}}
\put(1416.0,801.0){\rule[-0.200pt]{4.818pt}{0.400pt}}
\put(220.0,877.0){\rule[-0.200pt]{4.818pt}{0.400pt}}
\put(198,877){\makebox(0,0)[r]{1.8}}
\put(1416.0,877.0){\rule[-0.200pt]{4.818pt}{0.400pt}}
\put(220.0,113.0){\rule[-0.200pt]{0.400pt}{4.818pt}}
\put(220,68){\makebox(0,0){0}}
\put(220.0,857.0){\rule[-0.200pt]{0.400pt}{4.818pt}}
\put(342.0,113.0){\rule[-0.200pt]{0.400pt}{4.818pt}}
\put(342,68){\makebox(0,0){5}}
\put(342.0,857.0){\rule[-0.200pt]{0.400pt}{4.818pt}}
\put(463.0,113.0){\rule[-0.200pt]{0.400pt}{4.818pt}}
\put(463,68){\makebox(0,0){10}}
\put(463.0,857.0){\rule[-0.200pt]{0.400pt}{4.818pt}}
\put(585.0,113.0){\rule[-0.200pt]{0.400pt}{4.818pt}}
\put(585,68){\makebox(0,0){15}}
\put(585.0,857.0){\rule[-0.200pt]{0.400pt}{4.818pt}}
\put(706.0,113.0){\rule[-0.200pt]{0.400pt}{4.818pt}}
\put(706,68){\makebox(0,0){20}}
\put(706.0,857.0){\rule[-0.200pt]{0.400pt}{4.818pt}}
\put(828.0,113.0){\rule[-0.200pt]{0.400pt}{4.818pt}}
\put(828,68){\makebox(0,0){25}}
\put(828.0,857.0){\rule[-0.200pt]{0.400pt}{4.818pt}}
\put(950.0,113.0){\rule[-0.200pt]{0.400pt}{4.818pt}}
\put(950,68){\makebox(0,0){30}}
\put(950.0,857.0){\rule[-0.200pt]{0.400pt}{4.818pt}}
\put(1071.0,113.0){\rule[-0.200pt]{0.400pt}{4.818pt}}
\put(1071,68){\makebox(0,0){35}}
\put(1071.0,857.0){\rule[-0.200pt]{0.400pt}{4.818pt}}
\put(1193.0,113.0){\rule[-0.200pt]{0.400pt}{4.818pt}}
\put(1193,68){\makebox(0,0){40}}
\put(1193.0,857.0){\rule[-0.200pt]{0.400pt}{4.818pt}}
\put(1314.0,113.0){\rule[-0.200pt]{0.400pt}{4.818pt}}
\put(1314,68){\makebox(0,0){45}}
\put(1314.0,857.0){\rule[-0.200pt]{0.400pt}{4.818pt}}
\put(1436.0,113.0){\rule[-0.200pt]{0.400pt}{4.818pt}}
\put(1436,68){\makebox(0,0){50}}
\put(1436.0,857.0){\rule[-0.200pt]{0.400pt}{4.818pt}}
\put(220.0,113.0){\rule[-0.200pt]{292.934pt}{0.400pt}}
\put(1436.0,113.0){\rule[-0.200pt]{0.400pt}{184.048pt}}
\put(220.0,877.0){\rule[-0.200pt]{292.934pt}{0.400pt}}
\put(45,495){\makebox(0,0){$s_{t,b,\tau}$}}
\put(828,23){\makebox(0,0){$\tan \beta$}}
\put(1071,724){\makebox(0,0)[r]{$b$}}
\put(828,571){\makebox(0,0)[r]{$t$}}
\put(950,342){\makebox(0,0)[r]{$\tau$}}
\put(220.0,113.0){\rule[-0.200pt]{0.400pt}{184.048pt}}
\put(269,464){\usebox{\plotpoint}}
\multiput(269.00,464.59)(1.368,0.489){15}{\rule{1.167pt}{0.118pt}}
\multiput(269.00,463.17)(21.579,9.000){2}{\rule{0.583pt}{0.400pt}}
\multiput(293.00,473.59)(2.602,0.477){7}{\rule{2.020pt}{0.115pt}}
\multiput(293.00,472.17)(19.807,5.000){2}{\rule{1.010pt}{0.400pt}}
\multiput(317.00,478.61)(5.374,0.447){3}{\rule{3.433pt}{0.108pt}}
\multiput(317.00,477.17)(17.874,3.000){2}{\rule{1.717pt}{0.400pt}}
\put(342,481.17){\rule{4.900pt}{0.400pt}}
\multiput(342.00,480.17)(13.830,2.000){2}{\rule{2.450pt}{0.400pt}}
\put(390,482.67){\rule{6.023pt}{0.400pt}}
\multiput(390.00,482.17)(12.500,1.000){2}{\rule{3.011pt}{0.400pt}}
\put(366.0,483.0){\rule[-0.200pt]{5.782pt}{0.400pt}}
\put(463,482.67){\rule{6.023pt}{0.400pt}}
\multiput(463.00,483.17)(12.500,-1.000){2}{\rule{3.011pt}{0.400pt}}
\put(415.0,484.0){\rule[-0.200pt]{11.563pt}{0.400pt}}
\put(536,481.67){\rule{5.782pt}{0.400pt}}
\multiput(536.00,482.17)(12.000,-1.000){2}{\rule{2.891pt}{0.400pt}}
\put(488.0,483.0){\rule[-0.200pt]{11.563pt}{0.400pt}}
\put(585,480.67){\rule{5.782pt}{0.400pt}}
\multiput(585.00,481.17)(12.000,-1.000){2}{\rule{2.891pt}{0.400pt}}
\put(609,479.67){\rule{5.782pt}{0.400pt}}
\multiput(609.00,480.17)(12.000,-1.000){2}{\rule{2.891pt}{0.400pt}}
\put(560.0,482.0){\rule[-0.200pt]{6.022pt}{0.400pt}}
\put(658,478.67){\rule{5.782pt}{0.400pt}}
\multiput(658.00,479.17)(12.000,-1.000){2}{\rule{2.891pt}{0.400pt}}
\put(682,477.67){\rule{5.782pt}{0.400pt}}
\multiput(682.00,478.17)(12.000,-1.000){2}{\rule{2.891pt}{0.400pt}}
\put(633.0,480.0){\rule[-0.200pt]{6.022pt}{0.400pt}}
\put(731,476.67){\rule{5.782pt}{0.400pt}}
\multiput(731.00,477.17)(12.000,-1.000){2}{\rule{2.891pt}{0.400pt}}
\put(755,475.67){\rule{5.782pt}{0.400pt}}
\multiput(755.00,476.17)(12.000,-1.000){2}{\rule{2.891pt}{0.400pt}}
\put(779,474.67){\rule{6.023pt}{0.400pt}}
\multiput(779.00,475.17)(12.500,-1.000){2}{\rule{3.011pt}{0.400pt}}
\put(706.0,478.0){\rule[-0.200pt]{6.022pt}{0.400pt}}
\put(828,473.67){\rule{5.782pt}{0.400pt}}
\multiput(828.00,474.17)(12.000,-1.000){2}{\rule{2.891pt}{0.400pt}}
\put(852,472.67){\rule{6.023pt}{0.400pt}}
\multiput(852.00,473.17)(12.500,-1.000){2}{\rule{3.011pt}{0.400pt}}
\put(804.0,475.0){\rule[-0.200pt]{5.782pt}{0.400pt}}
\put(901,471.67){\rule{5.782pt}{0.400pt}}
\multiput(901.00,472.17)(12.000,-1.000){2}{\rule{2.891pt}{0.400pt}}
\put(925,470.67){\rule{6.023pt}{0.400pt}}
\multiput(925.00,471.17)(12.500,-1.000){2}{\rule{3.011pt}{0.400pt}}
\put(877.0,473.0){\rule[-0.200pt]{5.782pt}{0.400pt}}
\put(974,469.67){\rule{5.782pt}{0.400pt}}
\multiput(974.00,470.17)(12.000,-1.000){2}{\rule{2.891pt}{0.400pt}}
\put(950.0,471.0){\rule[-0.200pt]{5.782pt}{0.400pt}}
\put(1023,468.67){\rule{5.782pt}{0.400pt}}
\multiput(1023.00,469.17)(12.000,-1.000){2}{\rule{2.891pt}{0.400pt}}
\put(998.0,470.0){\rule[-0.200pt]{6.022pt}{0.400pt}}
\put(1071,467.67){\rule{6.023pt}{0.400pt}}
\multiput(1071.00,468.17)(12.500,-1.000){2}{\rule{3.011pt}{0.400pt}}
\put(1047.0,469.0){\rule[-0.200pt]{5.782pt}{0.400pt}}
\put(1144,466.67){\rule{5.782pt}{0.400pt}}
\multiput(1144.00,467.17)(12.000,-1.000){2}{\rule{2.891pt}{0.400pt}}
\put(1096.0,468.0){\rule[-0.200pt]{11.563pt}{0.400pt}}
\put(1387,466.67){\rule{6.023pt}{0.400pt}}
\multiput(1387.00,466.17)(12.500,1.000){2}{\rule{3.011pt}{0.400pt}}
\put(1168.0,467.0){\rule[-0.200pt]{52.757pt}{0.400pt}}
\put(1412.0,468.0){\rule[-0.200pt]{5.782pt}{0.400pt}}
\put(269,809){\usebox{\plotpoint}}
\put(317,807.17){\rule{5.100pt}{0.400pt}}
\multiput(317.00,808.17)(14.415,-2.000){2}{\rule{2.550pt}{0.400pt}}
\multiput(342.00,805.95)(5.151,-0.447){3}{\rule{3.300pt}{0.108pt}}
\multiput(342.00,806.17)(17.151,-3.000){2}{\rule{1.650pt}{0.400pt}}
\multiput(366.00,802.95)(5.151,-0.447){3}{\rule{3.300pt}{0.108pt}}
\multiput(366.00,803.17)(17.151,-3.000){2}{\rule{1.650pt}{0.400pt}}
\multiput(390.00,799.94)(3.552,-0.468){5}{\rule{2.600pt}{0.113pt}}
\multiput(390.00,800.17)(19.604,-4.000){2}{\rule{1.300pt}{0.400pt}}
\multiput(415.00,795.94)(3.406,-0.468){5}{\rule{2.500pt}{0.113pt}}
\multiput(415.00,796.17)(18.811,-4.000){2}{\rule{1.250pt}{0.400pt}}
\multiput(439.00,791.93)(2.602,-0.477){7}{\rule{2.020pt}{0.115pt}}
\multiput(439.00,792.17)(19.807,-5.000){2}{\rule{1.010pt}{0.400pt}}
\multiput(463.00,786.93)(2.714,-0.477){7}{\rule{2.100pt}{0.115pt}}
\multiput(463.00,787.17)(20.641,-5.000){2}{\rule{1.050pt}{0.400pt}}
\multiput(488.00,781.93)(2.118,-0.482){9}{\rule{1.700pt}{0.116pt}}
\multiput(488.00,782.17)(20.472,-6.000){2}{\rule{0.850pt}{0.400pt}}
\multiput(512.00,775.93)(2.118,-0.482){9}{\rule{1.700pt}{0.116pt}}
\multiput(512.00,776.17)(20.472,-6.000){2}{\rule{0.850pt}{0.400pt}}
\multiput(536.00,769.93)(2.118,-0.482){9}{\rule{1.700pt}{0.116pt}}
\multiput(536.00,770.17)(20.472,-6.000){2}{\rule{0.850pt}{0.400pt}}
\multiput(560.00,763.93)(1.865,-0.485){11}{\rule{1.529pt}{0.117pt}}
\multiput(560.00,764.17)(21.827,-7.000){2}{\rule{0.764pt}{0.400pt}}
\multiput(585.00,756.93)(1.789,-0.485){11}{\rule{1.471pt}{0.117pt}}
\multiput(585.00,757.17)(20.946,-7.000){2}{\rule{0.736pt}{0.400pt}}
\multiput(609.00,749.93)(1.550,-0.488){13}{\rule{1.300pt}{0.117pt}}
\multiput(609.00,750.17)(21.302,-8.000){2}{\rule{0.650pt}{0.400pt}}
\multiput(633.00,741.93)(1.865,-0.485){11}{\rule{1.529pt}{0.117pt}}
\multiput(633.00,742.17)(21.827,-7.000){2}{\rule{0.764pt}{0.400pt}}
\multiput(658.00,734.93)(1.550,-0.488){13}{\rule{1.300pt}{0.117pt}}
\multiput(658.00,735.17)(21.302,-8.000){2}{\rule{0.650pt}{0.400pt}}
\multiput(682.00,726.93)(1.368,-0.489){15}{\rule{1.167pt}{0.118pt}}
\multiput(682.00,727.17)(21.579,-9.000){2}{\rule{0.583pt}{0.400pt}}
\multiput(706.00,717.93)(1.616,-0.488){13}{\rule{1.350pt}{0.117pt}}
\multiput(706.00,718.17)(22.198,-8.000){2}{\rule{0.675pt}{0.400pt}}
\multiput(731.00,709.93)(1.368,-0.489){15}{\rule{1.167pt}{0.118pt}}
\multiput(731.00,710.17)(21.579,-9.000){2}{\rule{0.583pt}{0.400pt}}
\multiput(755.00,700.93)(1.368,-0.489){15}{\rule{1.167pt}{0.118pt}}
\multiput(755.00,701.17)(21.579,-9.000){2}{\rule{0.583pt}{0.400pt}}
\multiput(779.00,691.93)(1.427,-0.489){15}{\rule{1.211pt}{0.118pt}}
\multiput(779.00,692.17)(22.486,-9.000){2}{\rule{0.606pt}{0.400pt}}
\multiput(804.00,682.93)(1.368,-0.489){15}{\rule{1.167pt}{0.118pt}}
\multiput(804.00,683.17)(21.579,-9.000){2}{\rule{0.583pt}{0.400pt}}
\multiput(828.00,673.93)(1.368,-0.489){15}{\rule{1.167pt}{0.118pt}}
\multiput(828.00,674.17)(21.579,-9.000){2}{\rule{0.583pt}{0.400pt}}
\multiput(852.00,664.93)(1.427,-0.489){15}{\rule{1.211pt}{0.118pt}}
\multiput(852.00,665.17)(22.486,-9.000){2}{\rule{0.606pt}{0.400pt}}
\multiput(877.00,655.93)(1.368,-0.489){15}{\rule{1.167pt}{0.118pt}}
\multiput(877.00,656.17)(21.579,-9.000){2}{\rule{0.583pt}{0.400pt}}
\multiput(901.00,646.92)(1.225,-0.491){17}{\rule{1.060pt}{0.118pt}}
\multiput(901.00,647.17)(21.800,-10.000){2}{\rule{0.530pt}{0.400pt}}
\multiput(925.00,636.93)(1.427,-0.489){15}{\rule{1.211pt}{0.118pt}}
\multiput(925.00,637.17)(22.486,-9.000){2}{\rule{0.606pt}{0.400pt}}
\multiput(950.00,627.93)(1.368,-0.489){15}{\rule{1.167pt}{0.118pt}}
\multiput(950.00,628.17)(21.579,-9.000){2}{\rule{0.583pt}{0.400pt}}
\multiput(974.00,618.93)(1.368,-0.489){15}{\rule{1.167pt}{0.118pt}}
\multiput(974.00,619.17)(21.579,-9.000){2}{\rule{0.583pt}{0.400pt}}
\multiput(998.00,609.93)(1.427,-0.489){15}{\rule{1.211pt}{0.118pt}}
\multiput(998.00,610.17)(22.486,-9.000){2}{\rule{0.606pt}{0.400pt}}
\multiput(1023.00,600.93)(1.550,-0.488){13}{\rule{1.300pt}{0.117pt}}
\multiput(1023.00,601.17)(21.302,-8.000){2}{\rule{0.650pt}{0.400pt}}
\multiput(1047.00,592.93)(1.368,-0.489){15}{\rule{1.167pt}{0.118pt}}
\multiput(1047.00,593.17)(21.579,-9.000){2}{\rule{0.583pt}{0.400pt}}
\multiput(1071.00,583.93)(1.616,-0.488){13}{\rule{1.350pt}{0.117pt}}
\multiput(1071.00,584.17)(22.198,-8.000){2}{\rule{0.675pt}{0.400pt}}
\multiput(1096.00,575.93)(1.550,-0.488){13}{\rule{1.300pt}{0.117pt}}
\multiput(1096.00,576.17)(21.302,-8.000){2}{\rule{0.650pt}{0.400pt}}
\multiput(1120.00,567.93)(1.550,-0.488){13}{\rule{1.300pt}{0.117pt}}
\multiput(1120.00,568.17)(21.302,-8.000){2}{\rule{0.650pt}{0.400pt}}
\multiput(1144.00,559.93)(1.550,-0.488){13}{\rule{1.300pt}{0.117pt}}
\multiput(1144.00,560.17)(21.302,-8.000){2}{\rule{0.650pt}{0.400pt}}
\multiput(1168.00,551.93)(1.865,-0.485){11}{\rule{1.529pt}{0.117pt}}
\multiput(1168.00,552.17)(21.827,-7.000){2}{\rule{0.764pt}{0.400pt}}
\multiput(1193.00,544.93)(1.789,-0.485){11}{\rule{1.471pt}{0.117pt}}
\multiput(1193.00,545.17)(20.946,-7.000){2}{\rule{0.736pt}{0.400pt}}
\multiput(1217.00,537.93)(1.789,-0.485){11}{\rule{1.471pt}{0.117pt}}
\multiput(1217.00,538.17)(20.946,-7.000){2}{\rule{0.736pt}{0.400pt}}
\multiput(1241.00,530.93)(1.865,-0.485){11}{\rule{1.529pt}{0.117pt}}
\multiput(1241.00,531.17)(21.827,-7.000){2}{\rule{0.764pt}{0.400pt}}
\multiput(1266.00,523.93)(2.118,-0.482){9}{\rule{1.700pt}{0.116pt}}
\multiput(1266.00,524.17)(20.472,-6.000){2}{\rule{0.850pt}{0.400pt}}
\multiput(1290.00,517.93)(2.602,-0.477){7}{\rule{2.020pt}{0.115pt}}
\multiput(1290.00,518.17)(19.807,-5.000){2}{\rule{1.010pt}{0.400pt}}
\multiput(1314.00,512.93)(2.208,-0.482){9}{\rule{1.767pt}{0.116pt}}
\multiput(1314.00,513.17)(21.333,-6.000){2}{\rule{0.883pt}{0.400pt}}
\multiput(1339.00,506.93)(2.602,-0.477){7}{\rule{2.020pt}{0.115pt}}
\multiput(1339.00,507.17)(19.807,-5.000){2}{\rule{1.010pt}{0.400pt}}
\multiput(1363.00,501.94)(3.406,-0.468){5}{\rule{2.500pt}{0.113pt}}
\multiput(1363.00,502.17)(18.811,-4.000){2}{\rule{1.250pt}{0.400pt}}
\multiput(1387.00,497.93)(2.714,-0.477){7}{\rule{2.100pt}{0.115pt}}
\multiput(1387.00,498.17)(20.641,-5.000){2}{\rule{1.050pt}{0.400pt}}
\multiput(1412.00,492.95)(5.151,-0.447){3}{\rule{3.300pt}{0.108pt}}
\multiput(1412.00,493.17)(17.151,-3.000){2}{\rule{1.650pt}{0.400pt}}
\put(269.0,809.0){\rule[-0.200pt]{11.563pt}{0.400pt}}
\put(269,320){\usebox{\plotpoint}}
\put(269,318.67){\rule{5.782pt}{0.400pt}}
\multiput(269.00,319.17)(12.000,-1.000){2}{\rule{2.891pt}{0.400pt}}
\put(293,317.17){\rule{4.900pt}{0.400pt}}
\multiput(293.00,318.17)(13.830,-2.000){2}{\rule{2.450pt}{0.400pt}}
\put(317,315.67){\rule{6.023pt}{0.400pt}}
\multiput(317.00,316.17)(12.500,-1.000){2}{\rule{3.011pt}{0.400pt}}
\put(342,314.17){\rule{4.900pt}{0.400pt}}
\multiput(342.00,315.17)(13.830,-2.000){2}{\rule{2.450pt}{0.400pt}}
\put(366,312.17){\rule{4.900pt}{0.400pt}}
\multiput(366.00,313.17)(13.830,-2.000){2}{\rule{2.450pt}{0.400pt}}
\put(390,310.17){\rule{5.100pt}{0.400pt}}
\multiput(390.00,311.17)(14.415,-2.000){2}{\rule{2.550pt}{0.400pt}}
\multiput(415.00,308.95)(5.151,-0.447){3}{\rule{3.300pt}{0.108pt}}
\multiput(415.00,309.17)(17.151,-3.000){2}{\rule{1.650pt}{0.400pt}}
\put(439,305.17){\rule{4.900pt}{0.400pt}}
\multiput(439.00,306.17)(13.830,-2.000){2}{\rule{2.450pt}{0.400pt}}
\multiput(463.00,303.94)(3.552,-0.468){5}{\rule{2.600pt}{0.113pt}}
\multiput(463.00,304.17)(19.604,-4.000){2}{\rule{1.300pt}{0.400pt}}
\multiput(488.00,299.95)(5.151,-0.447){3}{\rule{3.300pt}{0.108pt}}
\multiput(488.00,300.17)(17.151,-3.000){2}{\rule{1.650pt}{0.400pt}}
\multiput(512.00,296.94)(3.406,-0.468){5}{\rule{2.500pt}{0.113pt}}
\multiput(512.00,297.17)(18.811,-4.000){2}{\rule{1.250pt}{0.400pt}}
\multiput(536.00,292.95)(5.151,-0.447){3}{\rule{3.300pt}{0.108pt}}
\multiput(536.00,293.17)(17.151,-3.000){2}{\rule{1.650pt}{0.400pt}}
\multiput(560.00,289.94)(3.552,-0.468){5}{\rule{2.600pt}{0.113pt}}
\multiput(560.00,290.17)(19.604,-4.000){2}{\rule{1.300pt}{0.400pt}}
\multiput(585.00,285.93)(2.602,-0.477){7}{\rule{2.020pt}{0.115pt}}
\multiput(585.00,286.17)(19.807,-5.000){2}{\rule{1.010pt}{0.400pt}}
\multiput(609.00,280.94)(3.406,-0.468){5}{\rule{2.500pt}{0.113pt}}
\multiput(609.00,281.17)(18.811,-4.000){2}{\rule{1.250pt}{0.400pt}}
\multiput(633.00,276.93)(2.714,-0.477){7}{\rule{2.100pt}{0.115pt}}
\multiput(633.00,277.17)(20.641,-5.000){2}{\rule{1.050pt}{0.400pt}}
\multiput(658.00,271.93)(2.602,-0.477){7}{\rule{2.020pt}{0.115pt}}
\multiput(658.00,272.17)(19.807,-5.000){2}{\rule{1.010pt}{0.400pt}}
\multiput(682.00,266.93)(2.602,-0.477){7}{\rule{2.020pt}{0.115pt}}
\multiput(682.00,267.17)(19.807,-5.000){2}{\rule{1.010pt}{0.400pt}}
\multiput(706.00,261.93)(2.714,-0.477){7}{\rule{2.100pt}{0.115pt}}
\multiput(706.00,262.17)(20.641,-5.000){2}{\rule{1.050pt}{0.400pt}}
\multiput(731.00,256.93)(2.602,-0.477){7}{\rule{2.020pt}{0.115pt}}
\multiput(731.00,257.17)(19.807,-5.000){2}{\rule{1.010pt}{0.400pt}}
\multiput(755.00,251.93)(2.602,-0.477){7}{\rule{2.020pt}{0.115pt}}
\multiput(755.00,252.17)(19.807,-5.000){2}{\rule{1.010pt}{0.400pt}}
\multiput(779.00,246.93)(2.714,-0.477){7}{\rule{2.100pt}{0.115pt}}
\multiput(779.00,247.17)(20.641,-5.000){2}{\rule{1.050pt}{0.400pt}}
\multiput(804.00,241.93)(2.118,-0.482){9}{\rule{1.700pt}{0.116pt}}
\multiput(804.00,242.17)(20.472,-6.000){2}{\rule{0.850pt}{0.400pt}}
\multiput(828.00,235.93)(2.602,-0.477){7}{\rule{2.020pt}{0.115pt}}
\multiput(828.00,236.17)(19.807,-5.000){2}{\rule{1.010pt}{0.400pt}}
\multiput(852.00,230.93)(2.208,-0.482){9}{\rule{1.767pt}{0.116pt}}
\multiput(852.00,231.17)(21.333,-6.000){2}{\rule{0.883pt}{0.400pt}}
\multiput(877.00,224.93)(2.602,-0.477){7}{\rule{2.020pt}{0.115pt}}
\multiput(877.00,225.17)(19.807,-5.000){2}{\rule{1.010pt}{0.400pt}}
\multiput(901.00,219.93)(2.118,-0.482){9}{\rule{1.700pt}{0.116pt}}
\multiput(901.00,220.17)(20.472,-6.000){2}{\rule{0.850pt}{0.400pt}}
\multiput(925.00,213.93)(2.714,-0.477){7}{\rule{2.100pt}{0.115pt}}
\multiput(925.00,214.17)(20.641,-5.000){2}{\rule{1.050pt}{0.400pt}}
\multiput(950.00,208.93)(2.118,-0.482){9}{\rule{1.700pt}{0.116pt}}
\multiput(950.00,209.17)(20.472,-6.000){2}{\rule{0.850pt}{0.400pt}}
\multiput(974.00,202.93)(2.602,-0.477){7}{\rule{2.020pt}{0.115pt}}
\multiput(974.00,203.17)(19.807,-5.000){2}{\rule{1.010pt}{0.400pt}}
\multiput(998.00,197.93)(2.208,-0.482){9}{\rule{1.767pt}{0.116pt}}
\multiput(998.00,198.17)(21.333,-6.000){2}{\rule{0.883pt}{0.400pt}}
\multiput(1023.00,191.93)(2.602,-0.477){7}{\rule{2.020pt}{0.115pt}}
\multiput(1023.00,192.17)(19.807,-5.000){2}{\rule{1.010pt}{0.400pt}}
\multiput(1047.00,186.93)(2.602,-0.477){7}{\rule{2.020pt}{0.115pt}}
\multiput(1047.00,187.17)(19.807,-5.000){2}{\rule{1.010pt}{0.400pt}}
\multiput(1071.00,181.93)(2.714,-0.477){7}{\rule{2.100pt}{0.115pt}}
\multiput(1071.00,182.17)(20.641,-5.000){2}{\rule{1.050pt}{0.400pt}}
\multiput(1096.00,176.93)(2.602,-0.477){7}{\rule{2.020pt}{0.115pt}}
\multiput(1096.00,177.17)(19.807,-5.000){2}{\rule{1.010pt}{0.400pt}}
\multiput(1120.00,171.93)(2.602,-0.477){7}{\rule{2.020pt}{0.115pt}}
\multiput(1120.00,172.17)(19.807,-5.000){2}{\rule{1.010pt}{0.400pt}}
\multiput(1144.00,166.93)(2.602,-0.477){7}{\rule{2.020pt}{0.115pt}}
\multiput(1144.00,167.17)(19.807,-5.000){2}{\rule{1.010pt}{0.400pt}}
\multiput(1168.00,161.94)(3.552,-0.468){5}{\rule{2.600pt}{0.113pt}}
\multiput(1168.00,162.17)(19.604,-4.000){2}{\rule{1.300pt}{0.400pt}}
\multiput(1193.00,157.94)(3.406,-0.468){5}{\rule{2.500pt}{0.113pt}}
\multiput(1193.00,158.17)(18.811,-4.000){2}{\rule{1.250pt}{0.400pt}}
\multiput(1217.00,153.94)(3.406,-0.468){5}{\rule{2.500pt}{0.113pt}}
\multiput(1217.00,154.17)(18.811,-4.000){2}{\rule{1.250pt}{0.400pt}}
\multiput(1241.00,149.94)(3.552,-0.468){5}{\rule{2.600pt}{0.113pt}}
\multiput(1241.00,150.17)(19.604,-4.000){2}{\rule{1.300pt}{0.400pt}}
\multiput(1266.00,145.94)(3.406,-0.468){5}{\rule{2.500pt}{0.113pt}}
\multiput(1266.00,146.17)(18.811,-4.000){2}{\rule{1.250pt}{0.400pt}}
\multiput(1290.00,141.95)(5.151,-0.447){3}{\rule{3.300pt}{0.108pt}}
\multiput(1290.00,142.17)(17.151,-3.000){2}{\rule{1.650pt}{0.400pt}}
\multiput(1314.00,138.94)(3.552,-0.468){5}{\rule{2.600pt}{0.113pt}}
\multiput(1314.00,139.17)(19.604,-4.000){2}{\rule{1.300pt}{0.400pt}}
\multiput(1339.00,134.95)(5.151,-0.447){3}{\rule{3.300pt}{0.108pt}}
\multiput(1339.00,135.17)(17.151,-3.000){2}{\rule{1.650pt}{0.400pt}}
\put(1363,131.17){\rule{4.900pt}{0.400pt}}
\multiput(1363.00,132.17)(13.830,-2.000){2}{\rule{2.450pt}{0.400pt}}
\multiput(1387.00,129.95)(5.374,-0.447){3}{\rule{3.433pt}{0.108pt}}
\multiput(1387.00,130.17)(17.874,-3.000){2}{\rule{1.717pt}{0.400pt}}
\put(1412,126.17){\rule{4.900pt}{0.400pt}}
\multiput(1412.00,127.17)(13.830,-2.000){2}{\rule{2.450pt}{0.400pt}}
\end{picture}

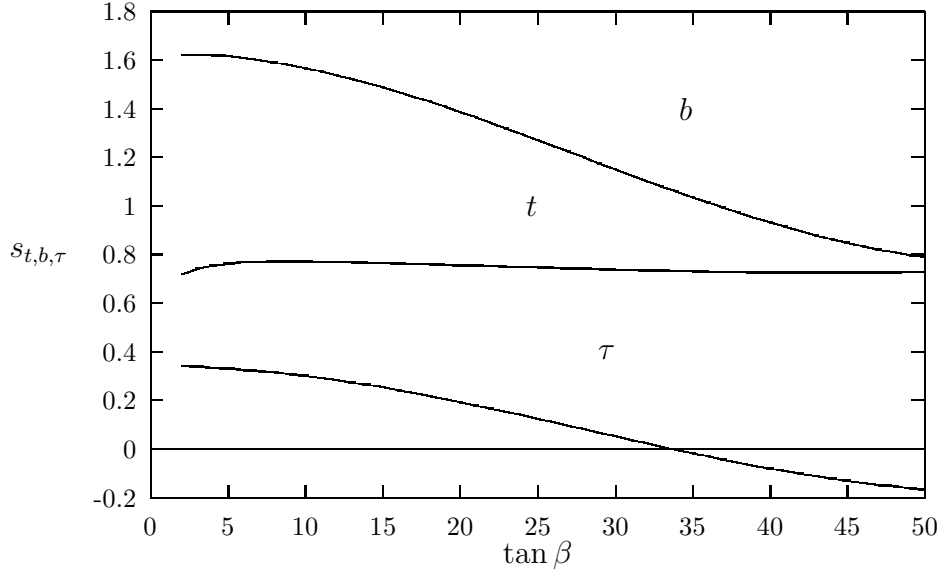
FIG. 3:$s_t, s_b, s_{\tau}$ against $\tan \beta$.
\end{center}

\noindent
We then require that the electroweak gauge symmetry
is correctly broken at $M_S$ to obtain
the sum rules,
\begin{eqnarray}
-\cos 2\beta ~m_{A}^{2}
&=&(s_{b}-s_{t} )M_{3}^{2} +2 (\hat{m}_{t}^{2}-
m_{t}^{2}) 
-2 (\hat{m}_{b}^{2}-
m_{b}^{2}) \nonumber\\
&=& (s_{\tau}-s_{t}) M_{3}^{2} +2 (\hat{m}_{t}^{2}-m_{t}^{2})
-2 (\hat{m}_{\tau}^{2}-m_{\tau}^{2}),
\label{sum}
\end{eqnarray}
where $m_{A}^{2}$ is the neutral pseudoscalar Higgs
mass squared, and $\hat{m}_{i}^{2}$
stands for the arithmetic mean of the two
corresponding scalar superparticle mass squared.
From the sum rules we also obtain
$(s_{b}-s_{\tau}) M_{3}^{2} \simeq 2 (\hat{m}_{b}^{2}-
\hat{m}_{\tau}^{2})$, which yields $
\hat{m}_{b}^{2} > \hat{m}_{\tau}^{2}$
because $s_{b}-s_{\tau} > 0$.
Note that $s_{\tau}$
becomes negative for $\tan\beta > 33$, which
gives a bound on $m_{H_1}^{2}$ for a given $M_3$,
$2 \hat{m}_{\tau}^{2}=s_{\tau} M_{3}^{2}
-(M_{Z}^{2}\cos 2\beta )/2
-m_{H_1}^{2} >0$.
If this is not satisfied, the $U(1)_{\rm EM}$ is broken.

The $s_{i}$'s are relatively stable 
against the deviation from 
the GUT scale sum rules (\ref{sum1}) and (\ref{sum2}),
which is shown in Fig.~4 for $m^2_{\Sigma(t)}/M_3^2$,
where we have varied the initial values at $M_{\rm GUT}$.
Fig.~4 shows a weak infrared attractiveness of
$m^2_{\Sigma(t)}/M_3^2$'s \cite{ross},
which is the reason of the stability.
So, the weak infrared attractiveness
works for  suppressing this uncertainty at $M_{\rm GUT}$,
but it is not strong
enough to wash out the information 
at at $M_{\rm GUT}$  \cite{polonsky}.

\begin{center}
\input{figmt350.tex}

FIG. 4:The evolution of
$m^2_{\Sigma(t)}/M_3^2$ for $\tan \beta =50$.
\end{center}

\noindent
We have also analyzed the infrared attractiveness 
\cite{ross} of $m^2_{\Sigma(b,\tau)}/M_3^2$ and 
$A_{t,b,\tau}$ for different values of 
$\tan\beta$ and found that
the infrared attractiveness is indeed
a general tendency \cite{ross}, 
but its degree differs among the
quantities and depends on $\tan\beta$.

As we have emphasized, the sum rules 
(\ref{sum}) are satisfied
under very general assumptions.
Before we close we summarize the most important ones.
The first one is  that the Yukawa couplings $Y_{ijk}$
in question are field-independent in $4D$ string models.
The next one is that below the string scale
the gauge-Yukawa unification is realized so that
the string inspired sum rules are
 RG invariant below the string scale
and are satisfied down to $M_{\rm GUT}$.
We do not need this assumption if
these sum rules
have a strong infrared attractiveness \cite{ross}
(which is model-dependent, of course).
It is then assumed that we have an $SU(5)$ type GUT
and below $M_{\rm GUT}$ the effective theory is the MSSM.

The sum rules could
be tested by future experiments, e.g., at LHC, and
experimental verifications of them would
give important hints on the nature of
unification and supersymmetry breaking.

\end{document}